\RequirePackage{lineno}
\documentclass[aps,prd,twocolumn,showpacs,superscriptaddress,groupedaddress]{revtex4}  
\usepackage{graphicx}  
\usepackage{dcolumn}   
\usepackage{bm}        
\usepackage{amssymb}   

\usepackage{booktabs}
\usepackage{longtable}
\usepackage{overpic}
\usepackage{multirow}
\usepackage{amsmath}
\usepackage{color}
\usepackage[colorlinks,linkcolor=blue,anchorcolor=blue,citecolor=blue]{hyperref}

\usepackage{subfigure}
\usepackage{tabularx}
\usepackage{gensymb}

\begin{document}



\title{Cross section measurement of $e^{+}e^{-} \to K_{S}^{0}K_{L}^{0}$ at $\sqrt{s}=2.00-3.08$~GeV}
\author{
\small
M.~Ablikim$^{1}$, M.~N.~Achasov$^{10,c}$, P.~Adlarson$^{64}$, S. ~Ahmed$^{15}$, M.~Albrecht$^{4}$, A.~Amoroso$^{63A,63C}$, Q.~An$^{60,48}$, ~Anita$^{21}$, X.~H.~Bai$^{54}$, Y.~Bai$^{47}$
% ...
\\
\vspace{0.2cm}
(BESIII Collaboration)\\
\vspace{0.2cm} {\it
$^{1}$ Institute of High Energy Physics, Beijing 100049, People's Republic of China\\
% ...
$^{4}$ Bochum Ruhr-University, D-44780 Bochum, Germany\\
% ...
$^{10}$ G.I. Budker Institute of Nuclear Physics SB RAS (BINP), Novosibirsk 630090, Russia\\
% ...
$^{15}$ Helmholtz Institute Mainz, Johann-Joachim-Becher-Weg 45, D-55099 Mainz, Germany\\
% ...
$^{21}$ Indian Institute of Technology Madras, Chennai 600036, India\\
% ...
$^{47}$ Southeast University, Nanjing 211100, People's Republic of China\\
$^{48}$ State Key Laboratory of Particle Detection and Electronics, Beijing 100049, Hefei 230026, People's Republic of China\\
% ...
$^{54}$ University of Jinan, Jinan 250022, People's Republic of China\\
% ...
$^{60}$ University of Science and Technology of China, Hefei 230026, People's Republic of China\\
% ...
$^{63}$ (A)University of Turin, I-10125, Turin, Italy; (B)University of Eastern Piedmont, I-15121, Alessandria, Italy; (C)INFN, I-10125, Turin, Italy\\
$^{64}$ Uppsala University, Box 516, SE-75120 Uppsala, Sweden\\
% ...
$^{c}$ Also at the Novosibirsk State University, Novosibirsk, 630090, Russia\\
% ...
}
\vspace{0.4cm}
}


\date{\today}

\begin{abstract}
  The cross sections of the process $e^{+}e^{-} \to
  K_{S}^{0}K_{L}^{0}$ are measured at fifteen center-of-mass energies
  $\sqrt{s}$ from $2.00$ to $3.08$~GeV with the BESIII detector at the Beijing Electron Positron Collider~(BEPCII).
  The results are found to be consistent with those obtained by BaBar. A resonant
  structure around $2.2$~GeV is observed, with a mass and width of
  $2273.7 \pm 5.7 \pm 19.3~{\rm MeV}/c^2$ and $86 \pm 44 \pm
  51~{\rm MeV}$, respectively, where the first uncertainties are
  statistical and the second ones are systematic. The product of its
  radiative width ($\Gamma_{e^+e^-}$) with its branching fraction to
  $K_{S}^{0}K_{L}^{0}$ ($Br_{K_{S}^{0}K_{L}^{0}}$) is $0.9 \pm 0.6 \pm
  0.7$~eV.

\end{abstract}
\pacs{13.60.Le, 13.66.Jn}
\maketitle

\section{\label{sec:level1}Introduction}

Among the light unflavored mesons, the strangeonium-like state
$\phi(2170)$ is particularly interesting. It was first reported in
$e^+ e^- \to \gamma_{\mathrm ISR} \phi f_0(980)$ by the BaBar
collaboration~\cite{Aubert:2006bu}, and then confirmed in $J/\psi \to
\eta \phi f_0(980)$ by the BESII collaboration~\cite{Ablikim:2007ab}
and in the $e^+ e^-\to \phi f_0(980)$ and $\phi \pi^+ \pi^-$ processes
by the Belle collaboration~\cite{Shen:2009zze}.  Subsequently, the
$\phi(2170)$ has been studied extensively by BaBar~\cite{Lees:2011zi,
  Aubert:2007ym, Aubert:2007ur, Aubert:2006bu},
Belle~\cite{Shen:2009zze}, BESII~\cite{Ablikim:2007ab}, and
BESIII~\cite{Ablikim:2014pfc, Ablikim:2019tpp, Ablikim:2020pgw,
  Ablikim:2020coo,Ablikim:2017auj,Ablikim:2020das,Ablikim:2018iyx,Ablikim:2020wyk}.

Initially, the strangeonium-like state $\phi(2170)$ was only observed
in hidden-strange decays, which makes its nature mysterious. Different
interpretations have been proposed.  In
Refs.~\cite{Wang:2006ri,Agaev:2019coa,Ke:2018evd,Dong:2020okt,Liu:2020lpw,Deng:2010zzd,Drenska:2008gr},
the $\phi(2170)$ is considered to be a tetraquark, while in
Refs.~\cite{Ding:2006ya, Ding:2007pc}, it is considered as an
$s\bar{s} g$ hybrid state.  Lattice QCD~\cite{Dudek:2011bn} and
QCD sum rule~\cite{Ho:2019org} investigations disfavor the $s\bar{s}
g$ hybrid interpretation.  Considering the near threshold location of
the $\phi(2170)$, various hadronic molecular possibilities have been
proposed, such as $\Lambda \bar{\Lambda}$
baryonium~\cite{Klempt:2007cp,Zhao:2013ffn,Yang:2019mzq}, a $\phi
K\bar{K}$~\cite{MartinezTorres:2008gy} or a $\phi
f_0(980)$~\cite{AlvarezRuso:2009xn} resonance. Besides these exotic
interpretations, the $\phi(2170)$ has been considered to be
conventional strangeonium, corresponding to
$3^3S_1$~\cite{Barnes:2002mu,Pang:2019ttv} or
$2^3D_1$~\cite{Ding:2007pc,Ding:2006ya,Wang:2012wa,Pang:2019ttv,Li:2020xzs}
states.  The predicted decay rates of $\phi(2170)\to K\bar{K}$ differ
among these theoretical interpretations.  For example, the branching
fraction is predicted to be $5\%-10\%$ under the assumption of a
$2^3D_1$ state~\cite{Wang:2012wa, Ding:2007pc} but close to zero in
the case of an $s\bar{s}g$ or $3^3Ss\bar{s}$~\cite{Ding:2006ya} state.
Therefore, an experimental measurement of the branching fraction of
$\phi(2170)\to K\bar{K}$ provides crucial information to distinguish
between the different interpretations.

Recently, the cross sections for $e^+ e^- \to K^+ K^-$ were measured
by the BESIII and BaBar
collaborations~\cite{Ablikim:2018iyx,BABAR:2019oes}.  A structure near
2.2~GeV was reported with a mass (width) differing from the world
averaged parameters of the $\phi(2170)$ by $3\sigma$~($2\sigma$).
The structure is not supported by Babar based on the measurements of the process $e^+e^-\to {K_{S}^{0}K_{L}^{0}}$~\cite{BABAR:2019oes}, though the uncertainties are very large which are more than $100\%$ in most of the energy intervals.
On the other hand, a theoretically guided fit to the BESIII cross sections for
$e^+e^-\to K^{+}K^{-}$ provided consistent results with respect to the
$\phi(2170)$ parameters~\cite{Chen:2020xho}. The structure observed in
the cross section measurements can also be explained as an
$\omega$-like state~\cite{Wang:2019jch}. In general, considering the
interferences between resonance and non-resonance contributions,
additional information from other processes, such as $e^+e^-\to
K_{S}^{0}K_{L}^{0}$, is needed. Although, this process has been
investigated in the past by the DM1~\cite{Mane:1980ep},
OLYa~\cite{Ivanov:1982cr},
CDM2~\cite{Akhmetshin:1999ym,Akhmetshin:2002vj,Akhmetshin:2003zn},
SND~\cite{Achasov:2000am,Achasov:2006bv} and
BaBar~\cite{BABAR:2019oes,Lees:2014xsh} collaborations, these
measurements mainly focused on the energy region below $2.0~$GeV.

In this work, we present Born cross section measurements of the
process $e^+e^-\to K_{S}^{0}K_{L}^{0}$.  The results obtained in the
overlapping center-of-mass region from $2.00-2.54$~GeV are compared to
previous measurements by BaBar~\cite{BABAR:2019oes}.  Moreover, we
present, for the first time, Born cross section measurements taken in
the interval from $2.54$ to $3.08$~GeV.  A fit is applied to the cross
section measurements of the $e^+e^-\to K_{S}^{0}K_{L}^{0}$ process,
and the resonant structure result is compared with that found by
BESIII~\cite{Ablikim:2018iyx} and BaBar~\cite{BABAR:2019oes} in
$e^+e^-\to K^{+}K^{-}$.

\section{\label{sec:level2}Detector and Monte Carlo Simulation}

The BESIII detector is a magnetic spectrometer~\cite{Ablikim:2009aa}
located at BEPCII~\cite{Yu:2016cof}. The cylindrical core of the
BESIII detector consists of a helium-based multilayer drift chamber
(MDC), a plastic scintillator time-of-flight system (TOF), and a
CsI(Tl) electromagnetic calorimeter (EMC), which are all enclosed in a
superconducting solenoidal magnet providing a 1.0~T magnetic
field. The solenoid is supported by an octagonal flux-return yoke with
resistive plate counter muon identifier modules interleaved with
steel. The acceptance of charged particles and photons is 93\% over
$4\pi$ solid angle. The charged-particle momentum resolution at
$1~{\rm GeV}/c$ is $0.5\%$, and the $dE/dx$ resolution is $6\%$ for
the electrons from Bhabha scattering. The EMC measures photon energies
with a resolution of $2.5\%$ ($5\%$) at $1$~GeV in the barrel (end
cap) region. The time resolution of the TOF barrel part is 68~ps,
while that of the end cap part is 110~ps.

The data samples used in this work are collected by the BESIII detector at fifteen center-of-mass~(c.m.) energies between $2.00$ and $3.08$~GeV
with an integrated luminosity of $583~{\rm pb}^{-1}$~\cite{Ablikim:2017wlt,Ablikim:2017duf}.

Monte Carlo (MC) samples simulated with a model of the complete
detector are used to determine detection efficiency, optimize event
selection criteria, and estimate backgrounds. Detector geometry,
material description, propagation and interactions with the detector
of the final-state particles are handled by {\sc
  GEANT4}-based~\cite{Agostinelli:2002hh} simulation software, {\sc
  BESIII Object Oriented Simulation Tool}~\cite{BOOSTBESIII}.

Signal and background samples are generated at each
c.m.~energy~($\sqrt{s}$). Signal MC samples of $e^+e^-\to
{K_{S}^{0}K_{L}^{0}}$ and $K_{S}^{0}\to\pi^+\pi^-$ are generated with
{\sc ConExc}~\cite{Ping:2013jka}. Non-hadronic backgrounds including continuum processes of $e^+e^-\to
e^+e^-$, $e^+e^-\to \gamma\gamma$ and $e^+e^-\to \mu^+\mu^-$ are
generated with {\sc Babayaga}~\cite{Balossini:2006wc}. Inclusive
hadronic samples~($e^+e^-\to q\bar{q}$) are generated with {\sc
  Luarlw}~\cite{Andersson:1999ui}.  Two-photon samples are generated
with {\sc BesTwoGam}~\cite{Nova:1990prh}.

\section{\label{sec:level3}Event selection and background analysis}
The momentum of the $K_{S}^{0}$ meson is reconstructed from its
$K_{S}^{0}\to\pi^+\pi^-$ decay. Events containing the reconstructed
$K_{S}^{0}$ candidates are retained for further analysis. The
$K_L^{0}$ meson is not detected directly; because of the two-body
decay, its presence is inferred by a requirement on the $K_{S}^{0}$
candidate momentum.  To select signal candidates, the following
criteria are applied:

\begin{itemize}

\item{} Exactly two oppositely-charged tracks are required without any
  requirement on neutral tracks. The distance of closest approach of the
  track with respect to the interaction point is required to be less
  than $20$~cm along the beam direction~($z-$axis of the BESIII
  coordinate system), while no requirement is made with respect to the
  transverse direction. Tracks are required to be within the
  acceptance of the MDC, $i.e.$~$|\cos \theta|<0.93$, where $\theta$
  is the polar angle between the track and the $z-$axis. A vertex fit
  is applied to constrain the two tracks to a common vertex, and
  subsequently a secondary vertex fit is performed to determine the
  flight distance $L$ and corresponding uncertainty $\delta L$, where
  $L$ corresponds to the separation between the secondary vertex and
  the interaction point.  We require $L/\delta L$ to be larger than
  $2$, as illustrated by the green vertical line in
  Figure~\ref{figLoverLerrInDifRegions}.  The invariant mass of the two
  tracks~($m_{\pi^+\pi^-}$), where the tracks are treated as $\pi^+$
  and $\pi^-$ candidates, is required to satisfy $|m_{\pi^+\pi^-} -
  m_{K_S^0}| < 35$~MeV$/c^2$, where $m_{K_S^0}$ is the mass of $K_S^0$
  taken from the Particle Data Group~(PDG)~\cite{Tanabashi:2018oca}. The
  signal yields are determined from fits to the invariant-mass
  distributions, as discussed in Section~\ref{sec:level4}.

\begin{figure}[pb]
\footnotesize
\begin{center}
\includegraphics[width=0.48\textwidth]{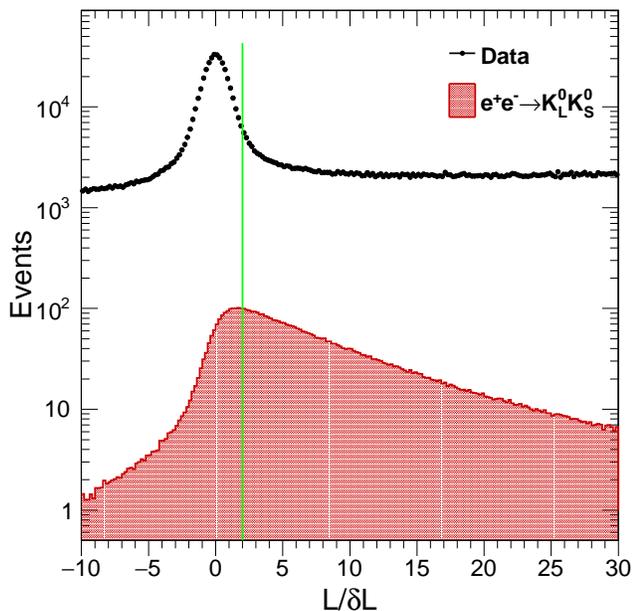}
\end{center}
\vspace*{-0.5cm}
\caption{$L/{\delta L}$ distribution for data taken at
  $\sqrt{s}=2.125$~GeV. Dots refer to data and the shaded area
  corresponds to simulated signal events normalized to the integrated
  luminosity of the data. The~(green)~vertical line indicates the
  requirement that is applied to select signal candidate events.}
\label{figLoverLerrInDifRegions}
\end{figure}

\item{} To reject backgrounds from the $e^+e^-\to e^+e^-$ and
  $e^+e^-\to \gamma\gamma$ processes, we require the ratio $E/cp$ between the
  deposited energy in the EMC ($E$) and the momentum measured by the
  MDC ($p$) to be less than $0.8$.

\item{} $|p_{\pi^+\pi^-} - p_{K_S^0}|<\sigma_p$ must be satisfied to
  suppress backgrounds from three (or more) body decays, where
  $p_{\pi^+\pi^-}$ is the momentum reconstructed from the $\pi^+\pi^-$ system, $p_{K_S^0}=\sqrt{\frac{s}{4} -
    ({m_{K_{S}^{0}}})^2}$ is the expected $K_S^0$ momentum, and
  $\sigma_p=15$~MeV$/c$ is the momentum resolution of the
  reconstructed $K_S^0$
  determined using the signal MC. The $p_{\pi^+\pi^-}$ distribution
  is shown in Figure~\ref{figKSp}.

\begin{figure}[pb]
\footnotesize
\begin{center}
\includegraphics[width=0.48\textwidth]{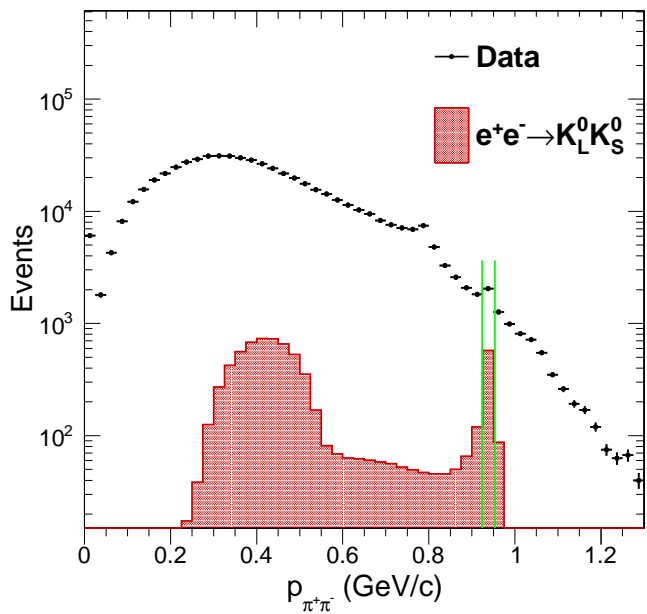}
\end{center}
\vspace*{-0.5cm}
\caption{$p_{\pi^+\pi^-}$ momentum distribution taken at
  $\sqrt{s}=2.125$~GeV. Dots refer to data and the shaded area depicts
  simulated signal events normalized to the integrated luminosity of
  the data. The wide peak on the left side in the simulated signal
  distribution stems from events that undergo initial-state radiation.
  The vertical lines indicate the window of the signal region.}
\label{figKSp}
\end{figure}

\end{itemize}

MC studies indicate that the non-hadronic background and two-photon process contribute less
than $5\%$ in the region $|m_{\pi^+\pi^-} - m_{K_S^0}| <
3\sigma_{K_S^0}$ at low c.m.~energies~($<2.396~{\rm GeV}$), where
$\sigma_{K_S^0}=4$~MeV$/c^2$ is the mass resolution of the pion pair
determined by fitting the signal MC shape, and it dominates at
$3.080$~GeV with a contribution of less than $20\%$.  No peaking
backgrounds were found from non-hadronic processes after applying the
previously described criteria at all c.m.~energies.  Detailed event type analysis with a generic tool, TopoAna~\cite{Zhou:2020ksj}, shows that the dominant
hadronic background channels are $e^+e^- \to K_S^0K_L^0\pi^0$, $e^+e^-
\to \pi^+\pi^-\pi^+\pi^-$, $e^+e^- \to \pi^+\pi^-\pi^0$ and $e^+e^-
\to (\gamma)\pi^+\pi^-$. A study using exclusive hadronic MC samples
showed that only at $3.080$~GeV a peaking background can be expected
from the process $e^+e^- \to K_S^0K_L^0\pi^0$. This will be further
discussed in Section~\ref{sec:level5}.

\section{\label{sec:level4}Cross section}
Born cross sections~($\sigma_B$) are obtained at each energy point by:
\begin{equation}
\begin{aligned}
\sigma_{B}=\frac{N_{sig}}{\epsilon(1+\delta)\mathcal{L}}\,,
\end{aligned}
\label{forXS}
\end{equation}

\noindent where $N_{sig}$ is the signal yield, $\epsilon$ is the
detection efficiency, $1+\delta$ is the correction factor including
vacuum polarization~(VP) and initial-state radiation~(ISR) effects,
and $\mathcal{L}$ is the integrated luminosity measured using
large-angle Bhabha scattering events with the method elucidated in
Ref.~\cite{Ablikim:2017wlt}. The branching ratio of
$K_S^0\to\pi^+\pi^-$ has been incorporated into $\epsilon$.

\begin{figure}[htpb]
\footnotesize
\centering
\includegraphics[width=0.48\textwidth]{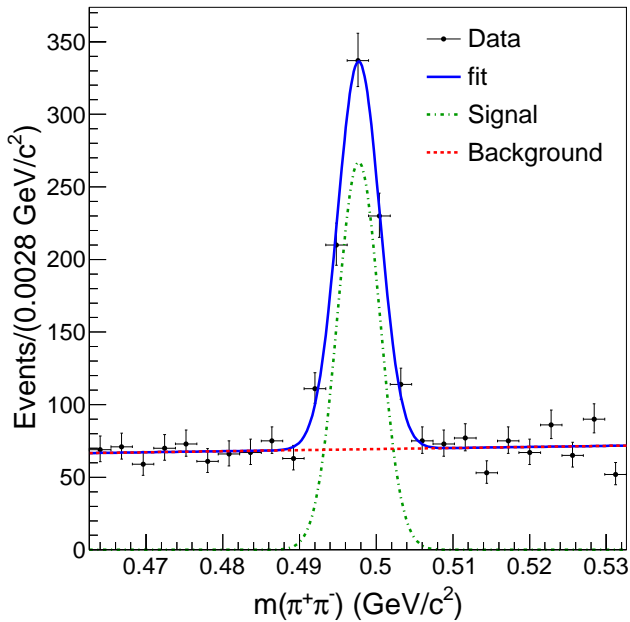}
\caption{$m_{\pi^+\pi^-}$ distribution of data taken at
  $\sqrt{s}=2.125$~GeV. The solid curve denotes the best fit
  through the data of the complete model, whereby the dash-dotted and
  dashed lines are the corresponding signal and background components,
  respectively.}
\label{figKSmfit}
\end{figure}

The signal yields are determined with an unbinned maximum-likelilood
fit to the invariant-mass distribution of $\pi^+\pi^-$ pairs of the
selected events obtained for each c.m.~energy point, where the signal
shape is described by a Gaussian function and the background is
represented with a zero-order Chebychev polynomial.  The fit range is
taken with a window of more than $8\sigma_{K_S^0}$ around the signal $K_S^0$.
  The mass and the
width of the Gaussian function is fixed to $m_{K_S^0}$ and
$\sigma_{K_S^0}$ for all the c.m.~energy points, except for the two
energies with the highest statistics~($2.000$ and $2.125$~GeV).  The
signal and background yields are set free for all c.m.~energies. As an
example, Figure~\ref{figKSmfit} illustrates the $m_{\pi^+\pi^-}$
distribution together with the corresponding fit result for data taken
at $\sqrt{s}=2.125$~GeV.

Both $\epsilon$ and $1+\delta$ depend on the line shape of the cross
sections and are determined via an iterative procedure.  In the first
iteration, the cross sections from $2.00~{\rm GeV}$ to $3.08~{\rm
GeV}$ are obtained and taken as initial inputs.
The cross sections below $2.00~{\rm GeV}$ are provided by previous
experiments~\cite{Mane:1980ep, Ivanov:1982cr, Akhmetshin:1999ym,
  Achasov:2000am, Akhmetshin:2002vj, Akhmetshin:2003zn,
  Achasov:2006bv, Lees:2014xsh} and fitted together with our
measurements above $2.00~{\rm GeV}$. The parameters $\epsilon$ and
$1+\delta$ are calculated according to the fit curve at each
c.m. energy and are taken as input for the next
iteration. The procedure is repeated until the measured Born cross
sections converge.

Results are summarized in Table~\ref{tabXSStat} with both statistical
and systematic uncertainties given in the last column.  The systematic
uncertainties are discussed in Section~\ref{sec:level5}.

\begin{table*}
  {\footnotesize{\caption{Born cross sections of the $e^+e^-\to K_S^0K_L^0$ process. The columns $N_{sig}$ and $N_{bkg}$ show
        the numbers of signal and background events determined by fitting the $m_{\pi^+\pi^-}$ distribution. The detection efficiency $\epsilon$,
        ISR and VP correction factor $1+\delta$, and the integrated luminosity $\mathcal{L}$ are summarized in the $4^{\rm th}$, $5^{\rm th}$, and $6^{\rm th}$
        column, respectively. The values presented in the column labeled with $\sigma_{B}$ correspond to the measured Born cross section,
        where the first uncertainty is statistical and the second one is systematic.}\label{tabXSStat}}}
\begin{ruledtabular}
\begin{tabular}{ccccccc} 
$\sqrt{s}$~(GeV)           & $N_{sig}$  & $N_{bkg}$   & $\epsilon(\times 10^{-4})$       &($1+\delta$)          & $\mathcal{L} ({\rm pb}^{-1})$    & $\sigma_{B}(\times 10^{-3}) ({\rm nb})$ \\ \hline 
2.000            & 185$\pm$18  & 341$\pm$22          & 541.2                                       & 6.09                   & 10.1                                           & 53.9$\pm$5.2$\pm$4.1             \\ 
2.050            & 51$\pm$9      & 115$\pm$12          & 448.8                                       & 7.48                    & 3.34                                          & 44.0$\pm$7.8$\pm$3.7             \\ 
2.100            & 101$\pm$13  & 252$\pm$18          & 289.6                                       & 11.77                  & 12.2                                          & 23.5$\pm$3.0$\pm$3.6             \\ 
2.125            & 658$\pm$34  & 1731$\pm$47        & 230.2                                       & 14.77                  & 108.                                          & 17.2$\pm$0.9$\pm$1.4             \\ 
2.150            & 14$\pm$6      & 101$\pm$11          & 198.1                                       & 16.85                   & 2.84                                         & 14.2$\pm$6.1$\pm$1.3             \\ 
2.175            & 67$\pm$10    & 125$\pm$13         & 213.0                                        & 15.59                   & 10.6                                         & 18.3$\pm$2.7$\pm$2.6             \\ 
2.200            & 81$\pm$11    & 146$\pm$14         & 266.9                                        & 12.13                   & 13.7                                         & 17.6$\pm$2.4$\pm$1.2             \\ 
2.232            & 98$\pm$12    & 133$\pm$13         & 360.9                                        &  9.03                    & 11.9                                         & 24.4$\pm$3.0$\pm$2.1             \\ 
2.309            & 116$\pm$13  & 171$\pm$15         & 259.4                                        &  13.04                  & 21.1                                         & 15.6$\pm$1.8$\pm$1.0             \\ 
2.386            & 27$\pm$7      & 78$\pm$10           & 82.0                                          & 40.84                   & 22.5                                         & 3.4$\pm$0.9$\pm$0.7                 \\ 
2.396            & 91$\pm$13    & 309$\pm$20         & 77.4                                          & 43.12                   & 66.9                                         & 3.9$\pm$0.6$\pm$0.5                 \\ 
2.644            & 52$\pm$9      & 90$\pm$11           & 51.8                                          & 59.69                    & 33.7                                        & 4.8$\pm$0.8$\pm$0.7               \\ 
2.646            & 57$\pm$9      & 70$\pm$10           & 51.8                                          & 59.30                    & 34.0                                        & 5.2$\pm$0.8$\pm$0.3               \\ 
2.900            & 43$\pm$9      & 91$\pm$11           & 42.9                                          & 68.07                    & 105.                                        & 1.4$\pm$0.3$\pm$0.2               \\ 
3.080            & 42$\pm$8      & 85$\pm$11           & 34.7                                          & 77.79                    & 126.                                        & 1.3$\pm$0.2$\pm$0.2               \\ 
\end{tabular}
\end{ruledtabular}
\end{table*}

\section{\label{sec:level5}Systematic uncertainties and line shape}
\subsection{\label{sec:level_1} Systematic uncertainties of the Born cross sections}
Several sources of the systematic uncertainties are estimated at each
c.m.~energy point, including uncertainties in the determination of the
$K_{S}^{0}$ selection efficiency, in applying the $E/cp$ requirement,
in the ISR and VP correction factors, in the integrated luminosity,
and in the fit procedure that was used to determine the signal
yield. The uncertainty in the $K_S^0\rightarrow \pi^+\pi^-$ branching
ratio is only $0.07\%$~\cite{Tanabashi:2018oca}, which is considered
to be negligible in this study.

The systematic uncertainty of the $K_{S}^{0}$ selection efficiency is
obtained using the control samples $J/\psi\to K^{*}(892)^{\mp}K^{\pm},
K^{*}(892)^{\mp}\to K_{S}^{0}\pi^{\mp}$ and $J/\psi\to\phi
K_{S}^{0}K^{\mp}\pi^{\pm}$, and the uncertainties are between
$2.2\%$ and $4.8\%$ depending on the reconstructed $K_S^0$
momentum~\cite{Ablikim:2012sf}. The uncertainty from the $E/cp$
requirement is estimated by changing the momentum $p$ of each track to
its value before applying the secondary-vertex fit.  The uncertainties
of the signal model, background model and fit range determine the
uncertainties of the signal yields. The uncertainty from the signal
model is estimated by changing the signal model to the shape predicted
by the MC data. The uncertainty due to the background model is
determined by replacing the background function with a first-order
Chebychev polynomial. The uncertainty associated to the fit range is
estimated by enlarging or reducing the fit range with an amount
corresponding to the mass resolution.

The systematic uncertainty of $\epsilon\times(1+\delta)$ is obtained
by fluctuating randomly all the fit parameters within the iteration
procedure by one $\sigma$ and taking into account the correlations
among the parameters. The distribution of the randomly produced
$\epsilon\times(1+\delta)$ is fitted by a Gaussian function, and the
width of the fitted parameter is defined as the systematic uncertainty of
$\epsilon\times(1+\delta)$. The uncertainty due to the luminosity is
estimated using large-angle Bhabha scattering events, which is about
$0.9\%$~\cite{Ablikim:2017duf,Ablikim:2017wlt}.

A MC study shows a peaking background from the process
$K_{S}^{0}K_{L}^{0}\pi^{0}$ at a center-of-mass energy of
$3.080$~GeV. However, the contribution normalized according to the
integrated data luminosity is expected to be only $2.6$ events. To
compensate for a possible incomplete simulation, such as an incorrect
angular distribution, the systematic uncertainty from the possible
$K_{S}^{0}K_{L}^{0}\pi^{0}$ background is increased to $3.1\%$
assuming the background level might be higher by $50\%$.

All the systematic uncertainties are listed in
Table~\ref{tabXSSyst}. The total systematic uncertainty is obtained by
summing the individual contributions in quadrature.

\begin{table*}
  {\footnotesize{\caption{The relative systematic uncertainties~(in
        $\%$) from the $K_S^0$ selection~($\epsilon(K_{S}^{0})$),
        $E/cp$, the ISR and VP correction factor~($1+\delta$), the
        luminosity~($\mathcal{L}$) and the fit on the invariant mass
        of $\pi^+\pi^-$ pair~($\rm {Fit}$).  The column $\rm {peak}$
        denotes the source from the peaking background and it has been
        estimated only at the c.m. energy of $3.080$~GeV as elucidated
        in the text. The total systematic uncertainty~($\rm {syst.}$)
        is calculated by summing the individual contributions in
        quadrature.  The relative statistical uncertainty~($\rm
        {stat.}$) is shown in the last column.}\label{tabXSSyst}}}
\begin{ruledtabular}
\begin{tabular}{ccccccccc}
$\sqrt{s}$~(GeV & $\epsilon(K_{S}^{0})$       & $E/cp$           & $\epsilon(1+\delta)$            & $\mathcal{L}$             & $\rm {Fit}$   & $\rm {peak}$    & $\rm {syst.}$  & $\rm {stat.}$ \\ \hline 
2.000            &2.99                                       & 0.53                & 0.63                                           &0.89                              & 6.87             &--                      & 7.6                & 9.7 \\ 
2.050            & 3.02                                      & 0.01                & 0.42                                           & 0.90                               & 7.74             &--                    & 8.4                 & 17.7 \\ 
2.100            & 2.92                                      & 0.01                 & 0.52                                          & 0.89                             & 15.14            &--                     & 15.5                     &  12.9  \\ 
2.125            & 2.82                                      & 0.15                & 0.67                                           & 0.69                             & 7.54               &--                    & 8.1                      &  5.2  \\ 
2.150            & 2.82                                      & 0.03              & 0.82                                            & 0.89                             & 8.93               &--                    & 9.4                  &  42.9\\ 
2.1750            & 3.47                                      & 0.03                & 0.65                                          & 0.90                             & 13.47             &--                    & 13.9                   &  14.9\\ 
2.200            &3.47                                       & 1.24                & 0.52                                          & 0.89                             & 5.42               &--                    & 6.6                      &  13.6 \\ 
2.232            &4.12                                       &  0.02                 & 0.72                                         & 0.90                             & 7.63               &--                    & 8.7                       &  12.2\\ 
2.309            & 3.17                                      & 0.01                &0.94                                          &  0.89                            & 5.24               &--                    & 6.2                     &  11.2\\ 
2.386            & 2.23                                      & 0.04                & 1.02                                         & 0.90                             & 20.65            &--                    & 20.8                  &  25.9\\ 
2.396            & 3.51                                      & 0.03                & 0.95                                         & 0.89                             &13.25             &--                     & 13.7                    &  14.3\\ 
2.644            & 3.38                                      & 1.93                 & 0.03                                         & 0.89                             & 14.60            &--                    & 15.1                   &  17.3\\ 
2.646            &3.38                                       & 1.75                & 0.03                                         & 0.89                             & 3.81              &--                    & 5.5                    &  15.8\\ 
2.900            & 2.63                                      & 2.33                & 0.04                                         & 0.89                             & 12.98            &--                    & 13.5                  &  20.9\\ 
3.080            & 4.8                                        & 2.38               & 0.04                                         & 0.84                             & 14.75            & 3.1                 & 15.7                  &  19.1\\ 
\end{tabular}
\end{ruledtabular}
\end{table*}

\subsection{\label{sec:level_2} Line shape}

\begin{figure}[htpb]
\footnotesize
\centering
\includegraphics[width=0.48\textwidth]{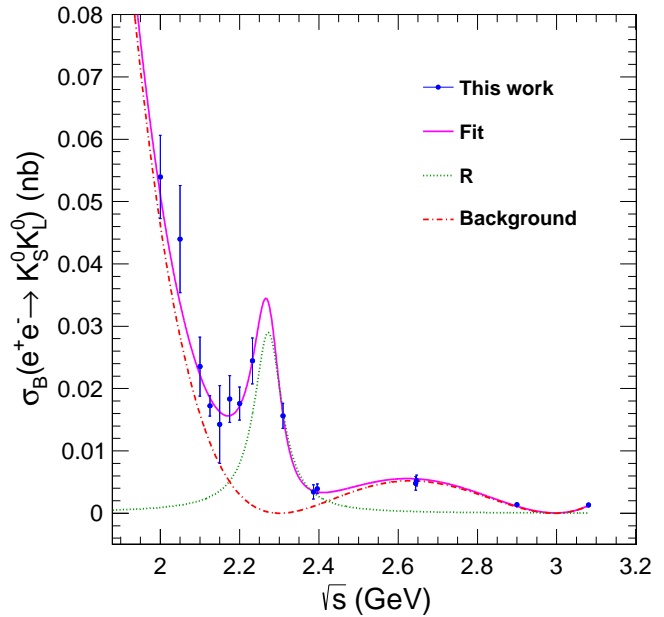}
\vspace*{-8pt}
\caption{Line shape of the process $e^+e^-\to K_{S}^{0}K_{L}^{0}$ and fit
  curves. Points are data, solid curve shows the fit result, the
  dotted curve denotes the signal component and the dash-dotted line
  is the polynomial contribution.}
\label{figXSFitFinalwith26GeV}
\end{figure}

The line shape of the Born cross section of $e^+e^-\to
K_{S}^{0}K_{L}^{0}$, obtained from the results given in
Table~\ref{tabXSStat}, is displayed in
Figure~\ref{figXSFitFinalwith26GeV}. A resonance structure $R$ around
$2.2$~GeV is observed. The cross section data are fitted by

\begin{equation}
\begin{aligned}
\sigma_{B}=&\frac{M^2\beta(s)^3}{s\beta(M^2)^3} |\sqrt{\sigma} BW(s)+P(s)e^{i\phi}|^2\,,
\end{aligned}
\label{forLineShape}
\end{equation}
\newline

\noindent where $\beta(s)=\sqrt{1-4m_{K_S^0}^2/s}$; $s$ is the square
of the c.m.~energy; $BW(s)=M\Gamma/(M^2-s-i\sqrt{s}\Gamma)$ is a
Breit-Wigner function describing the resonance; $M$, $\Gamma$ and
$\sigma$ are the mass, width and peak cross section of the resonance,
respectively; $P(s)=c_{p_0}+c_{p_1} {\sqrt {s}} + c_{p_2} s$ is a
second-order polynomial function that is used to describe the
nonresonant contribution, $c_{p_i}$ corresponds to the coefficient of
the $i^{th}$-degree polynomial function, and $\phi$ is the relative
phase between nonresonant and resonant amplitudes.

The least-squares~($\chi^2$) method is used to perform the fit with
both statistical and systematic uncertainties taken into account.  The
$\chi^2$ is obtained via a matrix~(see Eq.~(1) in
Ref.~\cite{Mo:2003cna} and Eq.~(2) in Ref.~\cite{Mo:2006bea}) in which
correlation effects of the various terms are included. Uncertainties
from the $K_S^0$-selection efficiency, $1+\delta$, luminosity and
$\epsilon$ are considered to be correlated, while the remaining ones
are treated as uncorrelated. The line shape and the individual
contributions obtained from the fit are shown in
Figure~\ref{figXSFitFinalwith26GeV}.

The mass and width of the structure determined by the fit are $M =
2273.7 \pm 5.7$~MeV$/c^2$ and $\Gamma = 86 \pm 44$~MeV,
respectively, where the uncertainties are statistical.  The goodness
of the fit is $\chi^2/NDF=4.6/8$, and the statistical significance of
the structure is $7.5\sigma$.

Various sources of systematic uncertainties of the observed structure
are considered including those associated with the choice of the model
used to describe the nonresonant component, the description of its
width and the chosen fit range. To estimate the systematic
uncertainties, we changed the description of the nonresonant component
to a coherent sum of a second-order polynomial and continuum functions

\begin{equation}
\begin{aligned}
P(s)=&P^{\prime}(s)e^{i\phi} + c_c(\sqrt{s})^{\alpha}  e^{i \phi_c}\,,
\end{aligned}
\label{forP}
\end{equation}
\newline

\noindent where $P^{\prime}(s)$ and $\phi$ are the same as those
defined in Eq.~(\ref{forLineShape}) but only used in the fit when
$\sqrt{s} < c_{p_2}$, $c_c$ is the coefficient of the continuum
function and $\phi_c$ is the relative phase between continuum and
resonant amplitudes. The differences in the values of the peak cross
section, mass, and width with respect to the nominal ones are $\Delta
\sigma=0.0150$~$\rm nb$, $\Delta m=17.7$~${\rm MeV}$$/c^2$, and
$\Delta\Gamma=8.4$~MeV, respectively. By replacing the description of
the width with an energy dependent one~($\Gamma(s, m)=\Gamma\times
\frac{s}{m_{R}^{2}}(\frac{\beta(s, m_K^0)}{\beta(m^2, m_K^0)})^3$) in
Eq.~(\ref{forLineShape}), the peak cross section, mass, and width
change by an amount of $\Delta \sigma=0.0001$~$\rm nb$, $\Delta
m=2.2$~MeV$/c^2$, and $\Delta \Gamma=0.3$~MeV, respectively.
Uncertainties from the fit range are estimated by excluding the point
at the c.m. energy of $2.00$~GeV or the one at $3.08$~GeV.  $\Delta
\sigma_1$ and $\Delta \sigma_2$~($\Delta m_1$ and $\Delta m_2$,
$\Delta \Gamma_1$ and $\Delta \Gamma_2$) denote the differences of the
peak cross sections~(masses and widths) obtained by fitting all energy
points with a fit excluding those two energy points. Systematic
uncertainties associated with the fit range on the mass and width are
subsequently estimated by $\sqrt{(\Delta \sigma_1)^2 + (\Delta
  \sigma_2)^2}=0.0030$~nb, $\sqrt{(\Delta m_1)^2 + (\Delta
  m_2)^2}=7.5$~MeV$/c^2$, and $\sqrt{(\Delta \Gamma_1)^2 + (\Delta
  \Gamma_2)^2}=50.2$~MeV. Total systematic uncertainties are obtained
by taking the quadratic sum of all the differences, which amount to
$0.0153$~nb, $19.3$~MeV$/c^2$, and $50.9$~MeV on the peak cross
section, mass, and width, respectively. Only the statistic uncertainty on
$\phi$ is considered.

$\Gamma_{e^+e^-}Br_{K_S^0K_L^0}$ of the resonance $R$ is
calculated from the peak cross section by making use of $\sigma_{R}={12\pi C \Gamma_{e^+e^-}Br_{K_S^0K_L^0}}/{(\Gamma M^2)}$~\cite{Lees:2014xsh}, where
$\sigma_{R}$ represents the peak cross section obtained through
Eq.~\ref{forLineShape}, $Br_{K_S^0K_L^0}$ is the branching fraction of $R\to K_S^0K_L^0$,  $\Gamma_{e^+e^-}$ is partial width of $R\to e^+e^-$, $M$ and $\Gamma$ are the mass and width of the resonance, and
$C=0.3894~\times~10^{12}~{\rm nb}~{\rm
  MeV^2}/c^4$~\cite{Tanabashi:2018oca}. $\Gamma_{e^+e^-}Br_{K_S^0K_L^0}$
  for the process is obtained from the fit results and listed in
  Eq.~\ref{eqfitresults}. 

The $\chi^2$ obtained by the earlier-described matrix may cause a bias
in the fit~\cite{Mo:2003cna,Mo:2003jwa,Mo:2006bea,Mo:2007aea}. To
estimate the bias effect, an unbiased $\chi^2$ definition~(Eq.~(7) in
Ref.~\cite{Mo:2007aea}) is used to fit the line shape. The differences
between the two cases are negligible in this analysis.

The parameters of the resonance around $2.2$~GeV are 

\begin{equation}
\begin{aligned}
M=&2273.7 \pm 5.7 \pm 19.3~{\rm {MeV}}/c^2\,, \\
\Gamma=&86 \pm 44 \pm 51~{\rm {MeV}}\,, \\
\sigma=&0.0289\pm 0.0125\pm 0.0153~{\rm nb}\,, \\
\Gamma_{e^+e^-}Br_{K_S^0K_L^0}=&0.9\pm 0.6\pm 0.7~{\rm eV}\,, \\
\phi=&81.1\pm 17.4~{\rm deg} \,, \\
&{\rm or} -98.9\pm 23.0~{\rm deg}\,,
\end{aligned}
\label{eqfitresults}
\end{equation}
\newline
\noindent where the quoted uncertainties are statistical and
systematic, respectively. The mass and width are consistent within
$2\sigma$ with measurements of the mass and width of a similar
structure observed in $e^{+}e^{-} \to K^{+}K^{-}$ at
BESIII~\cite{Ablikim:2018iyx}, which gave $M=2239.2 \pm 7.1 \pm
11.3$~MeV$/c^2$ and $\Gamma=139.8 \pm 12.3 \pm 20.6$~MeV.

\section{\label{sec:level7} Summary}

\begin{figure}[pb]
\footnotesize
\begin{center}
\includegraphics[width=0.48\textwidth]{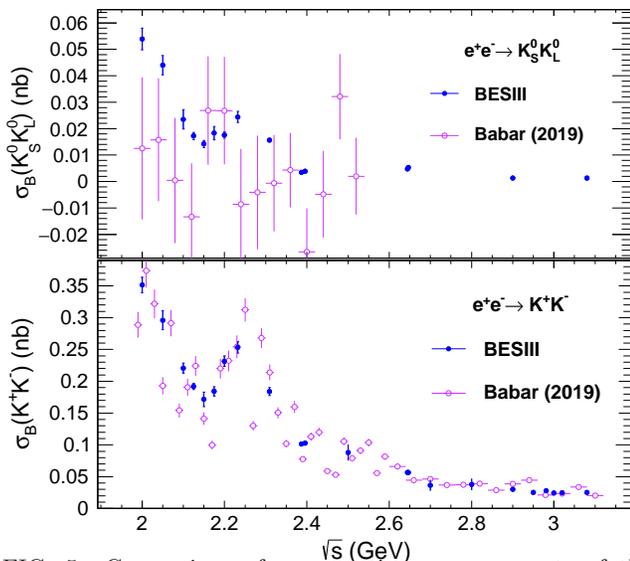}
\end{center}
\vspace*{-1cm}
\caption{Comparison of cross-section measurements of the processes
  $e^{+}e^{-} \to K_{S}^{0}K_{L}^{0}$ (top panel) and $e^{+}e^{-} \to
  K^{+}K^{-}$ (bottom panel) by BESIII (filled dots)~\cite{Ablikim:2018iyx} and BaBar (open
  circles)~\cite{BABAR:2019oes}. }
\label{figRatioKKKlKs}
\end{figure}

We report a measurement of the Born cross sections in $e^{+}e^{-} \to
K_{S}^{0}K_{L}^{0}$ from $\sqrt{s}=2.00$ to $3.08$~GeV obtained at
fifteen energy points with BESIII. The data are consistent within
$2\sigma$ with previous measurements by the BaBar
collaboration~\cite{BABAR:2019oes} in the overlap region from $2.00$
to $2.54$~GeV, but with a significantly improved precision as
demonstrated in Figure~\ref{figXSFitFinalwith26GeV}.  Moreover, the Born
cross sections from $2.54$ to $3.08$~GeV are reported for the first
time. A structure is observed around $2.2$~GeV, which is similar to
the one observed earlier in $e^+ e^- \to
K^+K^-$~\cite{Ablikim:2018iyx}. The results of both processes taken
with BESIII and BaBar are shown in Figure~\ref{figRatioKKKlKs} for
comparison.

A fit is applied to the data, where the mass and width of the
resonance are determined to be $M=2273.7 \pm 5.7 \pm 19.3$~MeV$/c^2$
and $\Gamma=86 \pm 44 \pm 51$~MeV, respectively. In addition,
$\Gamma_{e^+e^-}Br_{K_{S}^{0}K_{L}^{0}}$ is found to be $0.9\pm 0.6\pm
0.7~{\rm eV}$. The first uncertainties in the parameters are statistical and the second ones are systematic. The mass and width are consistent within $2\sigma$ and $1\sigma$, respectively, with the resonance parameters obtained by fitting the cross sections
for the process $e^+ e^- \to K^+ K^-$~($M=2239.2 \pm 7.1 \pm
11.3$~MeV$/c^2$ and $\Gamma=139.8 \pm 12.3 \pm 20.6$~MeV)~\cite{Ablikim:2018iyx}.

Comparing to one of the $1^{--}$ candidate of the structure $\phi(2170)$ by looking up the PDG~\cite{Tanabashi:2018oca}, the mass parameter obtained in this paper differs from the world average for more than $4\sigma$. The width parameter is consistent within $1\sigma$ compared to the world averaged width of $\phi(2170)$. The uncertainty on the width in this paper is large. 
For another $1^{--}$ candidate of the structure $\rho(2150)$~\cite{Tanabashi:2018oca}, the mass parameter in this paper is more than $5\sigma$ different from the world average and the wold averaged width is not given in the PDG~\cite{Tanabashi:2018oca}. But the mass and width are consistent with the individual  measurement of the process $e^+e^-\to\gamma\pi^+\pi^-$ by Babar~\cite{Lees:2012cj}. The conclusions support the discussions in the $e^+ e^- \to K^+ K^-$ study by BESIII~\cite{Ablikim:2018iyx}. Due to limit of the statistics, especially the cross section measurements above $2.4~\rm{GeV}$, it is difficult to discuss deeper on the structure found in this paper. More precise and fine interval measurements are needed.



\section*{Acknowledgements}
We appreciate Prof. Dianyong Chen from Southeast University~(SEU) for the important discussions and suggestions. The BESIII collaboration thanks the staff of BEPCII and the IHEP computing center and the supercomputing center of USTC for their strong support. This work is supported in part by National Key R$\&$D Program of China under Contracts Nos. 2020YFA0406400, 2020YFA0406300, 2015CB856700; National Natural Science Foundation of China (NSFC) under Contracts Nos. 11625523, 11635010, 11735014, 11822506, 11835012, 11935015, 11935016, 11935018, 11961141012; the Chinese Academy of Sciences (CAS) Large-Scale Scientific Facility Program; Joint Large-Scale Scientific Facility Funds of the NSFC and CAS under Contracts Nos. U1732263, U1832207 and U2032115; CAS Key Research Program of Frontier Sciences under Contracts Nos. QYZDJ-SSW-SLH003, QYZDJ-SSW-SLH040; 100 Talents Program of CAS; INPAC and Shanghai Key Laboratory for Particle Physics and Cosmology; ERC under Contract No. 758462; German Research Foundation DFG under Contracts Nos. Collaborative Research Center CRC 1044, FOR 2359; Istituto Nazionale di Fisica Nucleare, Italy; Ministry of Development of Turkey under Contract No. DPT2006K-120470; National Science and Technology fund; STFC (United Kingdom); The Knut and Alice Wallenberg Foundation (Sweden) under Contract No. 2016.0157; The Royal Society, UK under Contracts Nos. DH140054, DH160214; The Swedish Research Council; U. S. Department of Energy under Contracts Nos. DE-FG02-05ER41374, DE-SC-0012069. This paper is also supported by the NSFC under Contract No. 11605074, 11335008.

\end{document}